\documentclass[
showpacs,
floatfix,
aps,
prl,
twocolumn,
superscriptaddress,
amssymb,
]{revtex4-1}

\usepackage{times}
\usepackage{amssymb}
\usepackage{latexsym}
\usepackage[dvips]{graphicx}
\usepackage{amsmath}
\usepackage{graphicx}
\usepackage{dcolumn}
\usepackage{amsfonts}
\usepackage{bm}
\usepackage{epsfig}

\newcommand{\be}{\begin{equation}}
\newcommand{\ee}{\end{equation}}
\newcommand{\bea}{\begin{eqnarray}}
\newcommand{\eea}{\end{eqnarray}}

\def\bip{B_\parallel}

\def\nb{n_{\mathrm{3D}}}

\def\A{A_{R}}

\def\rxx{R_{xx}}
\def\ryy{R_{yy}}

\def\easy{\left < 110 \right >}
\def\hard{\left < 1\bar{1}0 \right >}
\def\x{\hat{x}}
\def\y{\hat{y}}

\def\bc{B_{\parallel}^c}

\def\a{\theta = 0^\circ}

\def\na{4.1}
\def\nb{4.3}
\def\ne{n_e}

\newcommand{\rfig}[1]{Fig.\,\ref{#1}}
\newcommand{\rFig}[1]{Figure \,\ref{#1}}

\newcommand{\rref}[1]{Ref.\,\onlinecite{#1}}

\begin{document}
\title{Quantum Hall stripes in high-density GaAs/AlGaAs quantum wells}
\author{X. Fu}
\affiliation{School of Physics and Astronomy, University of Minnesota, Minneapolis, Minnesota 55455, USA}
\author{Q. Shi}
\affiliation{School of Physics and Astronomy, University of Minnesota, Minneapolis, Minnesota 55455, USA}
\author{M. A. Zudov}
\email[Corresponding author: ]{zudov001@umn.edu}
\affiliation{School of Physics and Astronomy, University of Minnesota, Minneapolis, Minnesota 55455, USA}
\author{Y. J. Chung}
\affiliation{Department of Electrical Engineering, Princeton University, Princeton, New Jersey 08544, USA}
\author{K. W. Baldwin}
\affiliation{Department of Electrical Engineering, Princeton University, Princeton, New Jersey 08544, USA}
\author{L. N. Pfeiffer}
\affiliation{Department of Electrical Engineering, Princeton University, Princeton, New Jersey 08544, USA}
\author{K. W. West}
\affiliation{Department of Electrical Engineering, Princeton University, Princeton, New Jersey 08544, USA}
\begin{abstract}

We report on quantum Hall stripes (QHSs) formed in higher Landau levels of GaAs/AlGaAs quantum wells with high carrier density ($\ne > 4 \times 10^{11}$ cm$^{-2}$) which is expected to favor QHS orientation along unconventional $\hard$ crystal axis and along the in-plane magnetic field $\bip$. 
Surprisingly, we find that at $\bip = 0$ QHSs in our samples are aligned along $\easy$ direction and can be reoriented only perpendicular to $\bip$. 
These findings suggest that high density alone is not a decisive factor for either abnormal native QHS orientation or alignment with respect to $\bip$, while quantum confinement of the 2DEG likely plays an important role.
\end{abstract}
\received{2 October 2018; published 26 November 2018}
\maketitle

Electron nematic (or stripe) phases are known to form in a variety of condensed matter systems \cite{fradkin:2010,borzi:2007,daou:2010,chu:2010,okazaki:2011,falson:2018,hossain:2018}, including a two-dimensional electron gas (2DEG) in GaAs/AlGaAs quantum wells which offered the first realization of such broken symmetry states \cite{koulakov:1996,fogler:1996,lilly:1999a,du:1999,fradkin:1999,fradkin:2000}.
Arising from an interplay between exchange and direct Coulomb interactions \cite{koulakov:1996,fogler:1996}, quantum Hall stripes (QHSs) in a 2DEG are manifested by the resistivity minima (maxima) in the easy (hard) transport direction near half-integer filling factors, $\nu=9/2,11/2,13/2,...$\,. 
In a purely perpendicular magnetic field, QHSs in GaAs are nearly \cite{zhu:2002,pollanen:2015} always aligned along $\easy$ crystal direction, but the origin of such native symmetry-breaking potential remains a mystery \cite{sodemann:2013,koduvayur:2011,pollanen:2015}. 
Two experiments \citep{zhu:2002,cooper:2004}, however, have suggested that QHSs along $\hard$ direction are favored at higher carrier densities ($\ne \gtrsim 3 \times 10^{11}$ cm$^{-2}$), a regime which has not yet been systematically explored.

Shortly after the discovery of QHSs, it was realized that an in-plane magnetic field $\bip$ can easily reorient stripes \cite{lilly:1999b,pan:1999,cooper:2001} perpendicular to it.
This finding was well explained by theories considering the finite thickness of the 2DEG  \cite{jungwirth:1999,stanescu:2000}.
Subsequent experiments, however, revealed evidence for another mechanism which favors \emph{parallel} QHS alignment with respect to $\bip$ \cite{zhu:2002,zhu:2009,shi:2016c,shi:2017c}.
While the nature of this mechanism is not yet understood, experiments established that is it highly sensitive to both Landau and spin quantum numbers \citep{shi:2016c} and that it becomes increasingly important at higher electron densities \citep{shi:2017c}.
In particular, it was found that $\bip$, applied parallel to native QHSs at $\nu = 9/2$, could not alter their native orientation at all when $\ne > 3.5 \times 10^{11}$ cm$^{-2}$ \citep{shi:2017c}.
Unfortunately, densities above $\ne \approx 3.6 \times 10^{11}$ cm$^{-2}$ were not accessible because of the population of the second electrical subband.

Exploring QHSs in the regime of high carrier densities is interesting for several reasons. 
First, will native QHSs be oriented along $\easy$ or unconventional $\hard$ crystal axis as suggested by earlier studies \citep{zhu:2002,cooper:2004}?
If oriented along $\hard$, what would be the effect of $\bip$, e.g., will $\bip$ be able to alter orientation of such QHSs? 
In light of recent evidence that the mechanism favoring parallel-to-$\bip$ QHS alignment is itself anisotropic \citep{shi:2016c}, i.e., it appears sensitive to the direction of $\bip$ with respect to the crystal axes, answering this question may provide an insight not only on this mechanism but also on the native symmetry-breaking potential.

In this paper we investigate QHSs in high density ($\ne > 4 \times 10^{11}$ cm$^{-2}$) GaAs/AlGaAs quantum wells to determine (i) if QHSs are aligned along $\easy$ or $\hard$ crystal axis, and (ii) if QHSs can be reoriented by $\bip$, regardless of their initial alignment.
Our experiments reveal that our high-density samples exhibit well developed native QHSs with the orientation along conventional $\easy$ direction. 
In addition, we find that $\bip$ applied along native stripes produces a \emph{single} reorientation whereas $\bip$ applied perpendicular to QHSs does not alter their orientation.
We thus conclude that high $\ne$ alone is not a decisive factor for either abnormal native orientation of QHSs or their ultimate alignment with respect to $\bip$.
We suggest that quantum confinement is playing a crucial role in suppressing a symmetry-breaking mechanism which favors QHSs alignment along the in-plane magnetic field.


The 2DEG in sample A (B) resides in a GaAs quantum well of width 24 nm (25 nm) surrounded by Al$_{0.28}$Ga$_{0.72}$As barriers.
Sample A (B) utilized Si doping in narrow GaAs doping wells surrounded by thin Al$_{0.8}$Ga$_{0.2}$As layers and positioned at a setback distance of 73 nm (80 nm) on both sides of the GaAs well hosting the 2DEG.
After a brief low-temperature illumination, sample A (B) had the density $\ne \approx \na \times 10^{11}$ cm$^{-2}$ ($\ne \approx \nb \times 10^{11}$ cm$^{-2}$).
Low-temperature mobility was estimated to be $\mu \approx 1.2 \times 10^7$ cm$^2$V$^{-1}$s$^{-1}$ in sample A and $\mu \approx 0.9 \times 10^7$ cm$^2$V$^{-1}$s$^{-1}$ in sample B.
Both samples were $4\times 4$ mm squares with eight indium contacts fabricated at the corners and the midsides. 
The longitudinal resistances, $\rxx$ and $\ryy$, were measured at $T \approx 20$ mK using four-terminal, low-frequency lock-in technique.
An in-plane magnetic field (up to $\bip  = 16.7$ T) was introduced by tilting the sample about $\x \equiv \hard$ or $\y \equiv \easy$ axis, in two separate cooldowns.

\begin{figure}[t]
\includegraphics{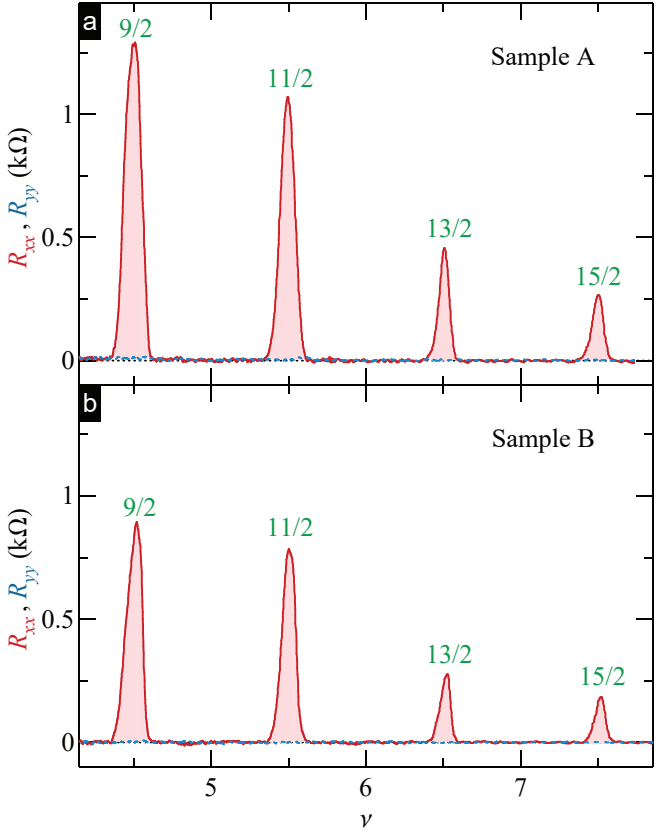}
\vspace{-0.1 in}
\caption{(Color online) 
$\rxx$ (solid line) and $\ryy$ (dotted line) versus filling factor $\nu$ measured in (a) sample A and (b) in sample B at $\bip =0$.
}
\vspace{-0.15 in}
\label{fig1}
\end{figure}
In \rfig{fig1}(a) and (b) we present $\rxx$ (solid line) and $\ryy$ (dotted line) measured in perpendicular magnetic field ($\bip = 0$) in sample A and B, respectively, as a function of the filling factor $\nu$ covering $N = 2$ and $N = 3$ Landau levels. 
Both data sets reveal formation of well-developed QHSs, as evidenced by sharp maxima in  $\rxx$ near $\nu = 9/2,11/2,13/2$, and $15/2$ and vanishing $\ryy$. 
Since $\rxx \gg \ryy$, we conclude that native QHSs are oriented along conventional $\y = \easy$ crystal axis.
Observation of conventional orientation in our samples with $\ne > 4 \times 10^{11}$ cm$^{-2}$ is somewhat surprising in light of previous experiments \citep{zhu:2002,cooper:2004} which indicated a transition from $\easy$ to $\hard$ native stripe orientation for densities above $3 \times 10^{11}$ cm$^{-2}$.
We thus conclude that high density alone is not a decisive factor for native QHS alignment along $\hard$ crystal axis.

\begin{figure}[t]
\centering
\includegraphics{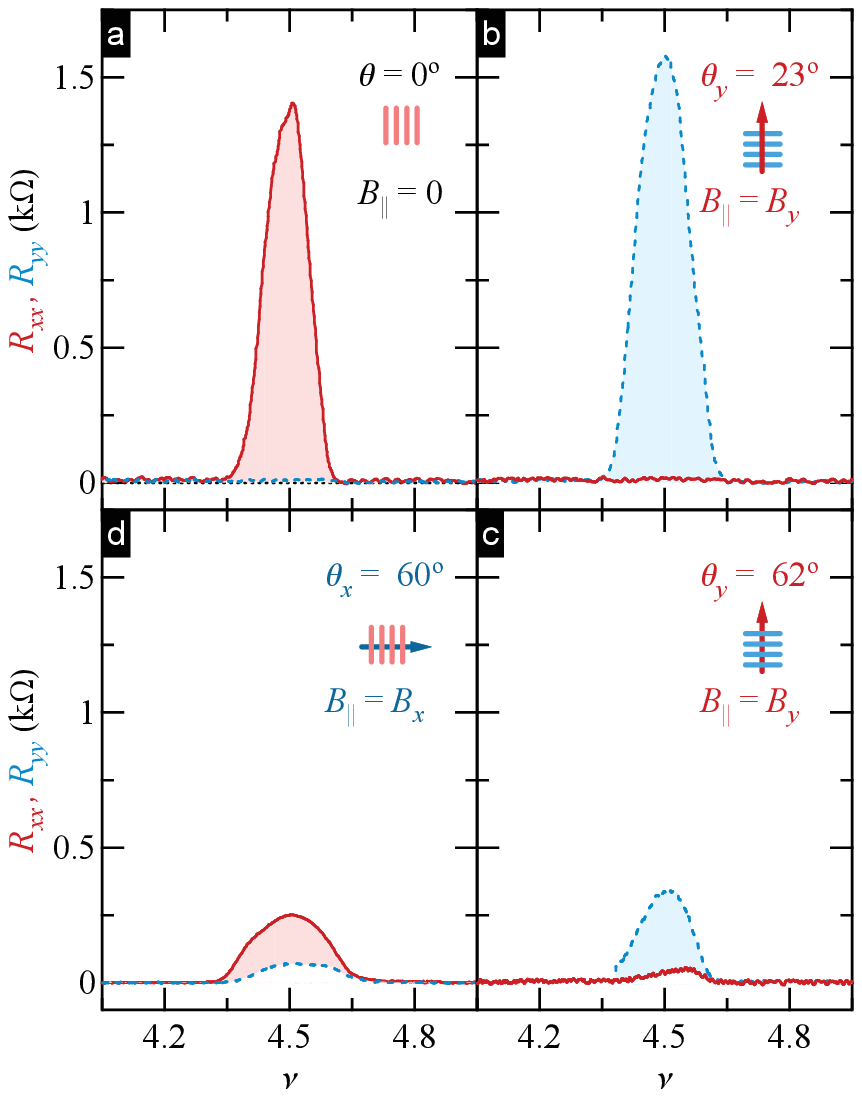}
\vspace{-0.1 in}
\caption{(Color online) 
$\rxx$ (solid line) and $\ryy$ (dotted line) versus $\nu$ in sample A at (a) $\a$, (b) $\bip = B_y$, $\theta_y = 23^{\circ}$, (c) $\bip = B_y$, $\theta_y = 62^{\circ}$, and (d) $\bip = B_x$, $\theta_x = 60^{\circ}$.
}
\vspace{-0.15 in}
\label{fig2}
\end{figure}

Having established native QHS orientation, we now turn to the effect of the in-plane magnetic field.
In \rfig{fig2} we present the results obtained in sample A near $\nu = 9/2$ with $\bip$ applied along either $\y$ or $\x$ direction.
\rFig{fig2}(a) shows the data at $\bip = 0$ revealing the native QHSs along $\y \equiv \easy$ direction ($\rxx \gg \ryy$).
As shown in \rfig{fig2}(b), when $\bip$ is applied parallel to the native stripes ($\bip = B_y$, $\theta_y = 23^{\circ}$), $\rxx$ and $\ryy$ switch places and we find $\rxx \ll \ryy$ indicating that stripes have been reoriented along $\x =\hard$-direction (perpendicular to $\bip$).
This reorientation is known since the discovery of the QHSs and has been observed in nearly every experiment examining the effect of $\bip = B_y$ \citep{lilly:1999b,pan:1999,cooper:2001,zhu:2009,pollanen:2015,shi:2016b,shi:2016c}.
However, observation of this reorientation in our high-density sample could not be readily anticipated since, as mentioned in the introduction, a recent study in a tunable-density 2DEG has found that the native QHS orientation remained unaffected by $\bip = B_y$, provided that the density is higher than $3.5 \times 10^{11}$ cm$^{-2}$ \citep{shi:2017c}.   

Upon further increase of $\bip = B_y$, stripes preserve their orientation along $\x = \hard$ direction remaining perpendicular to $\bip$ up to the highest field accessible in our experiment.
However, the resistance along hard (easy) axis eventually decreases (increases) as illustrated in \rfig{fig2}(c) showing the data at $\theta_y = 62^{\circ}$.
On the other hand, when $\bip$ is applied perpendicular to the native stripes ($\bip = B_x$), we observe no QHS reorientation up to the highest tilt angle.
As illustrated in \rfig{fig2}(d), at $\theta_x = 60^{\circ}$, $\rxx$ remains larger than $\ryy$ although the anisotropy ratio is greatly reduced, similar to what is observed in \rfig{fig2}(c) for $\theta_y = 62^{\circ}$.

\begin{figure}[t]
\includegraphics{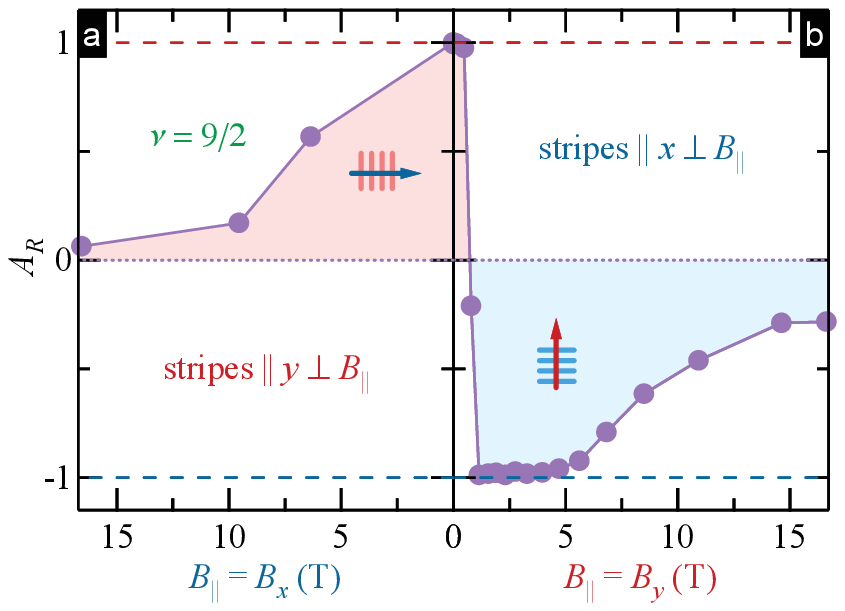}
\vspace{-0.1 in}
\caption{(Color online) 
Resistance anisotropy $\A \equiv (\rxx-\ryy)/(\rxx+\ryy)$ as a function of (a) $\bip = B_x$ and (b) $\bip = B_y$ at $\nu$ = 9/2.
}
\vspace{-0.15 in}
\label{fig3}
\end{figure}

To better illustrate the effect of $\bip$ observed in sample A at $\nu = 9/2$ we construct \rfig{fig3} which shows the resistance anisotropy $\A \equiv (\rxx-\ryy)/(\rxx+\ryy)$ as a function of (a) $\bip = B_x$ and (b) $\bip = B_y$.
With increasing $B_y$, $\A$ stays close to unity up to $B_y \approx 0.5$ T, vanishes at $B_y  = \bc \approx 0.8$ T, and reaches $\A \approx -1$ at $B_y \approx 1.1$ T.
The anisotropy then remains close to $-1$ up to $B_y \approx 4.9$ T after which $|\A|$ starts to decrease reaching $\A \approx -0.3$ at the highest $B_y \approx 16.7$ T [see \rfig{fig3}(b)].
As a function of $\bip = B_x$, $\A$ shows a decay and virtually vanishes at $B_x \approx 16.7$ T  [see \rfig{fig3}(a)] \citep{note:2}.

\begin{figure}[t]
\includegraphics{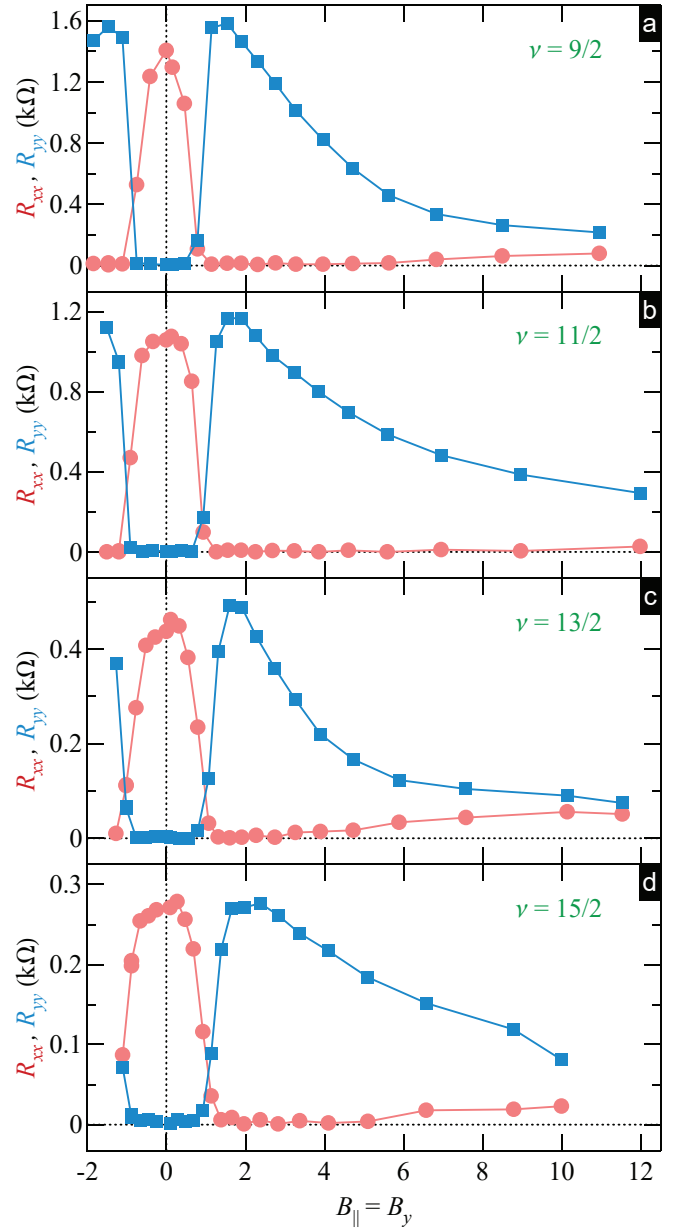}
\vspace{-0.1 in}
\caption{(Color online) 
$\rxx$ (circles) and $\ryy$ (squares) versus $\bip = B_y$ at (a) $\nu = 9/2$, (b) 11/2, (c) 13/2, and
(d) 15/2 measured in sample A. 
}
\vspace{-0.15 in}
\label{fig4}
\end{figure}

As we show next, at other half-integer filling factors in $N=2$ and $N=3$ Landau levels the response to $\bip = B_y$ is qualitatively the same, although there is some sensitivity to the spin index.
In \rfig{fig4} we present $\rxx$ (circles) and $\ryy$ (squares) versus $\bip = B_y$ at (a) $\nu = 9/2$, (b) 11/2, (c) 13/2, and (d) 15/2 measured in sample A. 
All data sets reveal one QHS reorientation occurring at $\bip = \bc$ which, consistent with the previous study \citep{shi:2016c}, monotonically increases with $\nu$ from $\bc \approx 0.8$ T at $\nu = 9/2$ to $\bc \approx 1.1$ T at $\nu = 15/2$.
It is also now clear that even though $A \approx -1$ at $1.1$ T $<B_y<$ $4.9$ T [see \rfig{fig3}(b)], the hard resistance $\ryy$ decreases by a factor of about three within these range at $\nu = 9/2$.
As $\ryy$ continues to drop with $B_y$, the decay of $|\A|$ observed at $B_y > 4.9$ T in \rfig{fig3}(b) occurs primarily due to the increase of $\rxx$ (which remained close to zero at $1.1$ T $<B_y<$ $4.9$ T).

Even though the data are qualitatively the same at all filling factors, closer examination reveals that the anisotropy ratio at lower spin branches ($\nu = 9/2$ and 13/2) decays noticeably faster with $\bip$ than at upper spin branches ($\nu = 11/2$ and 15/2).
While not well understood, sensitivity of the response to $\bip$ to the spin index has been noticed in previous experiments \citep{lilly:1999b,shi:2016c,shi:2017c,hossain:2018}.
Since the results obtained from sample B are essentially the same, the observed response of QHSs to $\bip$ in both of our high-density samples is similar to that reported by previous studies employing considerably lower density samples \citep{lilly:1999b,pan:1999,cooper:2001,shi:2017c}.

At the same time, the evolution of QHSs under applied $\bip$ observed in our high-density samples is qualitatively distinct from that seen in a tunable-density 30-nm quantum well in the higher density regime ($\ne \gtrsim 2.7 \times 10^{11}$ cm$^{-2}$) \citep{shi:2017c}.
The higher-density 2DEGs in our samples, however, have to reside in narrower quantum wells ($24-25$ nm) to avoid population of the second electrical subband.
It is therefore plausible that quantum confinement plays a crucial role in deciding the reorientation behavior. 

Since the reorientation under $\bip$ is believed to be due to finite thickness of the 2DEG, the effect of $\bip$ should become weaker in thinner 2DEGs.
In other words, everything else being equal, larger characteristic fields $B_y = \bc$ should be needed to reorient stripes in narrower quantum wells.
While $\bc$ can be affected by other factors, the obvious one being the native anisotropy energy, the observed values of $\bc$ in our experiment are in fact two to three times higher than those typically found in symmetric 30 nm quantum wells \citep{shi:2016c,shi:2017c}.
This finding is in agreement with \rref{pollanen:2015} which experimentally established strong sensitivity of $\bc$ to the separation between electrical subbands.

What is puzzling, however, is that a rather modest decrease of the quantum well width from 30 nm to 25 nm seems to a have a dramatic influence on the reorientation behavior; despite higher $\ne$ and much higher values of $\bip$ reached in our experiment, we find no range of $\bip$ which favors stripes parallel to $\bip$, in contrast to \rref{shi:2017c}.
This finding indicates that quantum confinement suppresses the mechanism responsible for parallel stripe alignment with respect to $\bip$ much more strongly than the one favoring perpendicular stripes.
This suppression seems to fully overwhelm any enhancement anticipated due to higher density \citep{shi:2017c}.

One should note that in the experiment which established that parallel-to-$\bip$ stripes are more likely to occur at higher carrier densities at a given $\nu$, the width of the 2DEG was increasing with $\ne$ as the quantum well became more symmetric under positive voltage applied to the backgate \citep{note:s}.
While complementary measurements of QHS orientations at $\nu = 9/2$ and 11/2 performed at fixed $B_z$ and $\bip = B_y$ seem to rule out the change of confinement as a primary driver of the transition to a parallel-to-$\bip$ QHS alignment, comparison of spin-up and spin-down branches might not be straightforward even when they belong to the same Landau level \citep{shi:2016c,hossain:2018}.

In summary, our experiments establish that electron density, while likely relevant, is not a decisive factor for either abnormal native orientation of QHSs or their ultimate alignment with respect to in-plane field. 
Instead, quantum confinement plays a crucial role in determining QHSs alignment with respect to $\bip$. 
In particular, we found that the recently identified mechanism which favors QHSs along $\bip$ is strongly suppressed in narrower 2DEGs, despite their considerably higher carrier density.
These finding should be useful for future theories aiming to explain what causes a particular QHSs alignment with respect to the in-plane magnetic field.
Understanding of the role of the in-plane field might also help to unveil the origin of the native QHS orientation, which remains a long-standing mystery despite continuing efforts.

\begin{acknowledgements}
We thank G. Jones, S. Hannas, T. Murphy, J. Park, A. Suslov, and A. Bangura for technical support.
The work at Minnesota was supported by the U.S. Department of Energy, Office of Science, Basic Energy Sciences, under Award \# ER 46640-SC0002567.
L.N.P. and K.W.W. of Princeton University acknowledge the Gordon and Betty Moore Foundation Grant No. GBMF 4420, and the National Science Foundation MRSEC Grant No. DMR-1420541.
A portion of this work was performed at the National High Magnetic Field Laboratory, which is supported by National Science Foundation Cooperative Agreement No. DMR-1644779 and the State of Florida.
\end{acknowledgements}


\end{document}